\documentclass[conference, letterpaper]{IEEEtran}
\IEEEoverridecommandlockouts

\usepackage[utf8]{inputenc}
\usepackage{pgfplots}

\usepackage[bibstyle=ieee,
    maxbibnames=12,
    sorting=none]{biblatex}

\DeclareBibliographyCategory{skipbib}
\newcommand{\skipbib}[1]{\addtocategory{skipbib}{#1}}

\usepackage{amsmath,amssymb,amsfonts}

\usepackage{algorithmic}
\usepackage{graphicx}
\usepackage{pgf}
\usepackage{tikz}
\usetikzlibrary{arrows,decorations.pathmorphing,decorations.pathreplacing,backgrounds,positioning,fit,matrix}
\usetikzlibrary{shapes,calc}

\usepackage{xcolor}
\definecolor{sron0}{HTML}{332288}
\definecolor{sron1}{HTML}{88CCEE}
\definecolor{sron2}{HTML}{117733}
\definecolor{sron3}{HTML}{DDCC77}
\definecolor{sron4}{HTML}{CC6677}

\usepackage{textcomp}
\usepackage{csquotes}
\usepackage{xcolor}
\ifCLASSOPTIONcompsoc \usepackage[caption=false,font=normalsize,labelfon
t=sf,textfont=sf]{subfig}
\else
\usepackage[caption=false,font=footnotesize]{subfig}
\fi

\usepackage{xspace}
\newcommand*{\eg}{e.g.\@\xspace}
\newcommand*{\ie}{i.e.\@\xspace}

\usepackage{tabularx}
\usepackage{booktabs}
\usepackage{balance}

\begin{document}

\title{Discharged Payment Channels:\\Quantifying the Lightning Network's Resilience to\\
Topology-Based Attacks}

\author{\IEEEauthorblockN{Elias Rohrer}
\IEEEauthorblockA{Distributed Security Infrastructures \\
\textit{Technical University of Berlin}\\
elias.rohrer@tu-berlin.de}
\and
\IEEEauthorblockN{Julian Malliaris}
\IEEEauthorblockA{Distributed Security Infrastructures \\
\textit{Technical University of Berlin}\\
julian.e.malliaris@campus.tu-berlin.de}
\and
\IEEEauthorblockN{Florian Tschorsch}
\IEEEauthorblockA{Distributed Security Infrastructures \\
\textit{Technical University of Berlin}\\
florian.tschorsch@tu-berlin.de}
}
\maketitle

\begin{abstract}
    The Lightning Network is the most widely
    used payment channel network (PCN) to date, making it an attractive attack
    surface for adversaries. In this paper, we analyze the Lightning Network's
    PCN topology and investigate its resilience towards random
    failures and targeted attacks. In particular, we introduce the notions of
    channel exhaustion and node isolation attacks
    and show that the Lightning Network is susceptible to these attacks.
    In a preliminary analysis, we confirm that the Lightning Network can be classified as
    a small-world and scale-free network.
    Based on these findings, we develop a series of strategies for targeted attacks and
    introduce metrics that allow us to quantify the adversary's advantage.
    Our results indicate that an attacker who is able to remove a certain number
    of nodes should follow a centrality-based strategy, while a
    resource-limited attacker who aims for high efficiency
    should employ a highest ranked minimum cut strategy.
\end{abstract}

\section{Introduction}\label{sec:intro}
As cryptocurrencies gain more real-world relevance,
they are currently faced with serious scalability issues~\cite{croman2016scalingblockchains}.
Payment channel networks (PCNs) address some of these issues by
processing transactions off-chain.
To this end, nodes open so-called payment channels,
which are secured by the blockchain.
They can be used to realize direct payments between channel endpoints
as well as multi-hop payments, even though no direct channel exists.
As a consequence, PCNs facilitate high transaction rates
that are not limited by the sheer number of processable transactions anymore,
but individually by the network's maximum flow capacity
for source-destination paths.

While there are different payment channel designs~\cite{miller17sprites,green17bolt,wattenhofer2015duplexmicropayment,lind16teechan,heilman17tumblebit},
the Bitcoin Lightning Network~\cite{poon2015bitcoin} is arguably
the most prevalent PCN implementation to date:
at the time of writing, it features a large, rapidly growing network
of around 2,500 public nodes and a total capacity of
more than 400\,BTC ($\approx$~1.4 million USD).
At the same time, this makes the Lightning Network an attractive attack surface for adversaries.

In this paper, we analyze the Lightning Network's PCN topology and
provide details on random and targeted attacks.
Moreover, we introduce the notion of \emph{exhaustion attacks}, \emph{node
isolation} attacks, and develop a series of adversarial strategies.
Most notably, we show that the Lightning Network is susceptible to these attacks
and can indeed be disrupted.

To this end, we apply methods from the field of
network theory~\cite{freeman1977set,boccaletti2006complex,massey1951kolmogorov},
to determine essential graph properties of the Lightning Network.
In particular, we evaluate whether the Lightning Network can be considered
a small world~\cite{humphries2008network}
and/or scale free network~\cite{clauset2009power}.
While the latter is known to tolerate random node failures very well,
the first exhibits short path lengths and high clustering coefficients
when compared to random graphs.
Both characteristics support the assessment of the Lightning Network's PCN topology
in terms of centralization, fault tolerance, and performance.

We use the insights from our network analysis
to investigate possible strategies for targeted attacks,
which may lead to a network split.
In particular, we not only consider denial-of-service (DoS) attacks,
but also discuss how malicious Lightning Network nodes can exhaust payment channels,
effectively isolating entire nodes.
An adversary can use this attack vector to render whole parts of the network unreachable.
We develop a series of attack strategies and quantify the adversary's advantage along a number of metrics,
including reachability, payment success, and maximum flow.

Our results indicate that the Lightning Network can be classified as
a small-world as well as scale-free network topology.
Accordingly, our assumption that the network is generally vulnerable to targeted attacks is confirmed.
We demonstrate the network's susceptibility to targeted attacks
by conducting an empirical analysis based on a recent snapshot of the Lightning Network.
The results suggest that an attacker who is able to remove a certain number
of nodes should follow a centrality-based strategy,
while a resource-limited attacker who aims for high efficiency should employ
a highest ranked minimum cut strategy.
The main contributions of our work can be summarized as follows:
\begin{itemize}
	\item We study the current state of the Lightning Network's PCN topology
	and assess its resilience to random failures and targeted attacks.
	\item We systematize topology-based attacks against PCNs.
	\item We introduce channel exhaustion and node isolation attacks as
	additional attack vectors.
	\item We develop various adversarial strategies
	and quantify their prospects in terms of the adversarial success.
\end{itemize}

The remainder of this paper is structured as follows.
Section~\ref{sec:analysis} is
concerned with our findings regarding the network's basic characteristics.
In Section~\ref{sec:attacks}, we present the possibility of
exhaustion attacks and derive adversarial strategies. Moreover, we
evaluate the feasibility of such attacks and quantify their prospects
and costs in Section~\ref{sec:evaluation}.
In Section~\ref{sec:relwork}, we discuss related work,
before concluding the paper in Section~\ref{sec:conclusion}.

\section{Lightning Network Analysis}\label{sec:analysis}
In the Lightning Network, every node has a global view of the payment channel network (PCN) topology
and is responsible for finding routes on the basis of this data set,
\ie, conducting source routing.
The routing information is distributed
by broadcasting \texttt{channel\_announcement} and \texttt{channel\_update} messages
via the Lightning Network's own peer-to-peer network.
These messages contain relevant channel information,
such as signatures, channel capacities, and fees.
However, mainly due to privacy reasons, actual channel balances are not included.
As a consequence, it cannot be predetermined
whether sufficient funding is available to route a payment.
In case of a failed payment attempt, the client needs to repeat the process
until successfully completed.

The payment channel design of the Lightning Network ensures balance security
for multihop payments through the use of hash time locked contracts (HTLCs).
In this construction, the receiver of a payment first issues a payment request
containing a hash $H(r)$ of a secret value $r$. The sender then initiates the
payment and sends a transaction to the first node on the payment path that may
only be redeemed by revealing the pre-image $r$. This process is
repeated for every hop on the path, until a transaction reaches the receiver
of the payment. The receiver then redeems the payment by publishing $r$, allowing all
nodes on the path to claim their locked funds as well.
To enable atomicity, HTLCs allow to revert stuck payment negotiations after the expiration of the transaction time locks,
which currently defaults to one hour~\cite{BOLT}.

We can see that HTLC processing requires all
participants to be online and respondent. Otherwise, if a node goes offline,
funds may be locked for an extended time period.
In the worst case, it is even possible that an adversary publishes outdated states,
effectively stealing coins.
The robustness of the PCN topology therefore is the baseline for the resilience of the Lightning Network,
\eg, to node failures due to DoS attacks.
In the following, we evaluate metrics like the betweenness centrality,
clustering coefficient, and degree distribution to draw conclusions on
whether the topology exhibits properties similar to small-world or scale-free networks.
These common properties provide insights on the degree of centralization and
the sensitivity to random failures as well as targeted attacks.

\begin{figure}
    \input{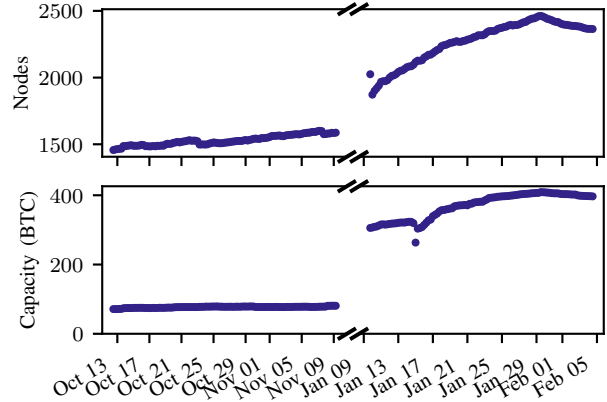}
    \vspace{-2em}
    \caption{Time series showing the Lightning Network's number of nodes and its total capacity
    (\ie, sum of all payment channel capacities).
    }
  \label{fig:timeseries}
\end{figure}

\subsection{Data Collection and Methodology}
Over the span of two measurement periods (Oct.--Nov. 2018
and Jan.--Feb. 2019), we gathered information on the Lightning Network's topology.
To this end, we used two virtual machines based on Ubuntu Server $18.04$,
which run \texttt{bitcoind}\footnote{\fullcite{bitcoind}}\skipbib{bitcoind}
and \texttt{lnd}\footnote{\fullcite{lnd}}\skipbib{lnd}, respectively.
We utilized the RPC \texttt{describegraph} to retrieve the topology
and regularly store snapshots in a \texttt{.json} file.
For the data analysis we developed a python evaluation script,
using the \texttt{networkx}~\cite{hagberg2008exploring}
and \texttt{powerlaw}~\cite{alstott2014powerlaw} libraries.
Our code and all data sets are available online\footnote{\url{https://gitlab.tu-berlin.de/rohrer/discharged-pc-data}}.

In general, a node's view of the topology depends on the information it gets
from its neighbors. Moreover, not all channels have to be announced publicly.
Therefore, there is no guarantee to have a complete view of the network.
However, we assume that the publicly available data characterizes the network's
essential traits.
While the network has immensely grown in terms of the number of nodes as
well as in total capacity the past months (see Figure~\ref{fig:timeseries}),
we can observe that the topology's characteristics have not changed significantly.
The following analysis is based on a recent data set, which was
captured on Feb.\ 1, 2019 0:00\,AM.

\subsection{Graph Measures and Metrics}

We consider the Lightning Network's PCN topology as a graph $G = (V,E)$,
where $V$ is the set of nodes and $E$ the set of edges, \ie, payment channels.
The degree $deg(v)$ of a vertex $v$ is defined as the number of its channels.
A channel between node $v_i$ and node $v_j$ is denoted as $e_{ij}$.
A path between two nodes consists of one or more channels.
The distance between two nodes is defined as the shortest path between these nodes.
The diameter is the longest distance between any two nodes in the network.
Similar to the diameter, the average path length is
defined as the average distance between any two nodes.

\emph{Betweenness} is a metric for centrality~\cite{freeman1977set}.
The betweenness of a node is the number of shortest paths
between any two nodes in the network that pass through the node.
The betweenness centrality $c_B(v)$ of a node $v$ is given by
\[c_B(v) = \sum_{s, t \in  V}\frac{\sigma(s,t|v)}{\sigma(st)},\]
where $\sigma(st)$ is the number of all shortest paths between $s$ and $t$,
and $\sigma(s,t|v)$ is the number of all shortest paths between $s$ and $t$
that include $v$.
To normalize the value such that $c_B \in [0,1]$,
$c_B$ is divided by the number of all pairs of nodes that do not include $v$,
that is, $(n-1)(n-2)/2$ for undirected graphs with $n$ being the total number of nodes.
Accordingly, it describes the share of shortest paths that pass this node.
In general, betweenness centrality is an indicator of how much control a node has over the network.

A subgraph $G' = (V',E')$ where $V'\subseteq V$ and $E'\subseteq E$
is called connected component if any node $v'\in V'$ can be
reached by any other node in $V'$.
A graph is biconnected if it is still connected after removing any arbitrary node.
A \emph{biconnected component} is a largest possible biconnected subgraph.
If a node is member of more than one biconnected component,
it is called an articulation point or cut vertex.
As the removal of nodes that have a high betweenness and/or are articulation
points may have an increased impact on network connectivity, they are
potential targets of directed attacks.

Graph metrics typically gain meaning only in
comparison to other graphs of the same size.
In Table~\ref{tab:comparison} we compare the PCN graph
to a random Erd\"{o}s-Renyi graph~\cite{erdds1959random}
and a scale-free Barabasi-Albert graph~\cite{barabasi1999emergence}.
Note that while the graphs share similar parameters,
the scale-free graph has less edges
due to the method of preferential attachment.

From the comparison, we can observe that currently all three graph types share a
diameter of 5--6 hops. At the same time, though, the PCN graph has the lowest
average distance,
which is a favorable property for users as it reduces the failure probability
and saves routing fees.
In terms of centrality, we compared the central point dominance,
defined as the maximum betweenness centrality of all nodes.
We can see that the PCN and the scale-free graph have a central point dominance
more than ten times as high as the random graph's.
The result suggests that the Lightning Network relies on few central nodes
in order to process payments.

\begin{table}
\begin{center}
\caption{Comparison of graph measures for different graph types.}
\label{tab:comparison}
  \begin{tabularx}{\columnwidth}{Xrrr}
    \toprule
     & PCN & Scale-free & Random \\
    \midrule
    Node count & 2400 & 2400 & 2400 \\
    Edge count & 13884 & 	11975&	13941\\
    Diameter & 	6 & 5 &	6		\\
    Average distance &  2.92 & 3.25 & 3.45 \\
    Central point dominance & 0.16 & 0.09 & 0.005\\
    \bottomrule
  \end{tabularx}
\end{center}
\end{table}

\subsection{Small-World Networks}

Small-world networks are characterized by nodes that tend to cluster
and have a high density of edges.
More formally, the diameter grows logarithmically with the number of nodes. %
In order to test the \enquote{small-world-ness} of the Lightning Network,
we use the method introduced in~\cite{humphries2008network}.
It is based on comparing the clustering coefficient
to a clustering coefficient of a random graph with similar parameters,
which serves as a reference.

While different definitions of the global clustering coefficient exist,
we use the \emph{transitivity} definition~\cite{boccaletti2006complex}.
Accordingly, the clustering coefficient $C$ is defined as
\begin{equation*}\label{eq:cluster}
C = \frac{3 \cdot \text{number of triangles}}{\text{number of paths of length 2}}.
\end{equation*}
Note, a factor of $3$ is used to compensate
that each triangle has three paths of length $2$.
Thus, $C=1$ for cliques.

Now, let $L_g$ and $L_{r}$ denote the mean shortest path length of
the PCN graph $G$ and a random Erd\"{o}s-Renyi graph $R$, respectively.
Likewise, let $C_g$ and $C_r$ describe the clustering coefficient.
In accordance with~\cite{humphries2008network}, we consider a network as \emph{small-world} if $S \gg 1$, where
\[S = \frac{\gamma_g}{\lambda_g}
  \quad\text{with}\quad \gamma_g = \frac{C_g}{C_r}
  \,\text{ and }\, \lambda_g = \frac{L_g}{L_r}.\]

Applying the described method to our empirical data yields
$C_g = 0.085$ and $L_g = 2.92$ for the PCN graph and
$C_r = 0.005$ and $L_r = 3.45$ for the random graph.
We can already see that the Lightning Network is more clustered
and yields on average shorter distances. We can conclude that the Lightning
Network can be classified as a small-world network, as $S = 19.439 \gg 1$.

\subsection{Scale-Free Networks}
\begin{figure}
\input{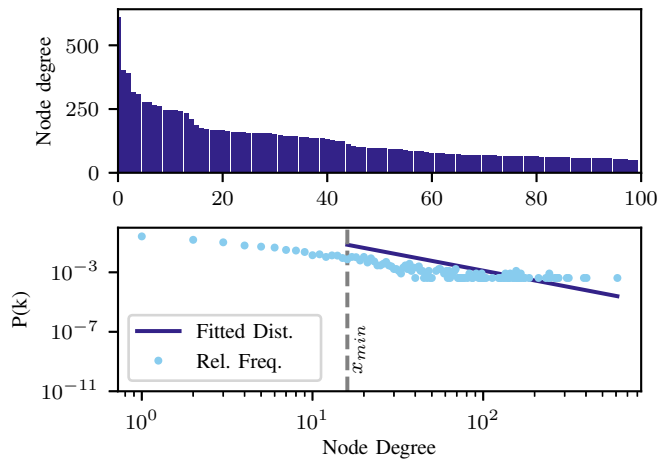}
  \caption{Degree distribution of the Lightning Network. The log-log plot
  additionally shows the fitted power-law distribution.}
  \label{fig:degree_distribution}
\end{figure}

Scale-free networks are characterized by a few nodes having a very high degree
and many nodes having low degrees.
More specifically, the degree distribution is similar to a power law distribution,
where the fraction of nodes $P(k)$ having a degree $k$ is described as
$P(k) \sim k^{-\alpha}$ with $\alpha$ typically ranging between 2 and 3~\cite{clauset2009power}.

Scale-free networks emerge if a new node can choose its neighbors freely
and prefers well-connected nodes.
In the Lightning Network, we have a comparable situation.
New nodes have an incentive to preferably open channels to highly connected nodes,
hence reaching a larger share of the network with fewer hops.
In comparison to random graphs, scale-free networks are generally robust to
random failures, as
the chance of a critical amount of high-degree nodes failing concurrently is very small.
However, since a few nodes have high degrees,
scale-free networks are prone to targeted attacks.

In Figure~\ref{fig:degree_distribution}, we show the degree distribution of the 100 nodes with highest degree in the Lightning
Network: the initial impression makes the hypothesis of a power-law
distribution, \ie, scale-free network, plausible.
To examine whether the Lightning Network is actually scale-free,
we investigate the degree distribution along the lines of~\cite{clauset2009power}.
The empirical data are plotted in a log-log plot in
Figure~\ref{fig:degree_distribution}.
On the x-axis we show the node degree $k$
and on the y-axis the probability for a certain node degree $P(k)$.
For a power law distribution we expect a negative linear trend,
where the slope determines the scaling factor $\alpha$.
However, this alone is not sufficient to draw conclusions.
To get sound results, we additionally perform a power-law fit
using a maximum likelihood estimator.
More specifically, we use the Kolmogorov-Smirnov (KS) distance~\cite{massey1951kolmogorov}
to determine the difference between the actual data and a proposed power-law fit.
By minimizing the KS distance for $x$, we retrieve an $x_{min}$.

With $\alpha$ and $x_{min}$, we can derive a power-law distribution,
but to draw conclusions it requires a goodness-of-fit test.
Based on a number of synthetic data sets and respectively fitted distribution parameters
(derived from the regression model),
it generates a $p$-value, which we use to accept or reject the hypothesis of a power-law distribution.
The authors of~\cite{clauset2009power} suggest to use
synthetic data sets with a high number of samples (ideally $10,000$ samples)
and to reject the scale-free hypothesis if $p \leq 0.1$.

Applying this method to the empirical PCN data yields
an $\alpha = 2.18$ and an $x_{min} = 16$.
For the goodness-of-fit test we created $10,000$ data sets, that in sum
yielded a $p$-value of $p=0.3314$, clearly substantiating the scale-free
hypothesis.

In conclusion, the degree distribution of the Lightning Network can
be classified as power-law distributed, suggesting a scale-free network
structure overall. Therefore, the network may benefit from the robustness property of scale-free networks against random failures.

\subsection{Robustness Analysis}
\begin{table}
\begin{center}
\caption{Average number of connected components after random failures
for differen graph types.}
\label{tab:remove_random}
  \begin{tabularx}{\columnwidth}{Xrrr}
    \toprule
   Failures & PCN & Scale-free & Random \\
   \midrule
   1  & 	1.13 & 	1.00 & 1.00	\\
   2  & 	1.20 & 	1.00 & 1.00	\\
   3  & 	1.61 & 	1.00 & 1.00	\\
   5  & 	2.86 & 	1.00 & 1.01	\\
   10 & 	3.66 & 	1.00 & 1.00	\\
   50 & 	14.00 & 1.00 & 1.03	\\
    \bottomrule
  \end{tabularx}
\end{center}
\end{table}

To draw conclusions about the robustness of the Lightning Network, we again compare
it to other graph types (Table~\ref{tab:remove_random}). We randomly removed a
certain amount of nodes to simulate random failures. Each time, the simulation
was run 100 times for the Lightning Network, a scale-free Barabasi-Albert
graph~\cite{barabasi1999emergence}, and a
random Erd\"{o}s-Renyi graph~\cite{erdds1959random}, respectively. The random removal of
nodes has nearly no impact on the random graph and the scale-free graph but
separates the PCN graph. The isolated components mostly consist of
one and two nodes and therefore will barely affect routing efficiency. Yet, the
graph separates and a random failure is very likely to separate at least one
node from the network.
In conclusion, the impact of random failures on the routing efficiency is very low.
Nevertheless, the prospects of targeted attacks seems promising.
Based on this insight, we discuss several attack vectors
and their impact on the network in the following sections.

\section{Attacking the Lightning Network}\label{sec:attacks}

As we have seen in the previous section, the Lightning Network is actually
rather centralized and exhibits a heavily skewed degree distribution.
This raises the question how the network topology copes with attacks
that target specific points of interest.
In the following, %
we present specific attack vectors, including channel exhaustion and node isolation attacks,
and discuss a number of feasible attack strategies. Moreover, we introduce
metrics that allow us to quantify the adversarial advantage of each strategy.

\subsection{Adversary Model}
With its growing success, the Lightning Network becomes an increasingly
interesting target for different kinds of adversaries. We assume an active
adversary that may participate in the peer-to-peer network, or attack
its topology from the outside.
At this point, we do not make any assumptions about the adversary's resources
as this will be a parameter of our evaluation.
In general, we assume however that the
adversary is always eager to act as efficiently as possible, \ie, minimizing
resources to maximize its adversarial advantage.

The adversary's motivation (or goal) may vary
and therefore determines the attack vectors and strategies.
For example, an adversary may be interested in eliminating
single nodes, \eg, to impede or censor their participation
in the network. She could also be interested in disrupting the
network as a
whole and aim for a partitioning attack that could impair
the payment processing or even inhibit it entirely.
Lastly, the adversary may be a \enquote{selfish} node in the Lightning Network,
\eg, a payment hub,
and interested in increasing her fee gain
by sabotaging competing nodes and payment paths.

\tikzset{node distance=1.3cm, every node/.style={thick, font=\small\sffamily},
routenode/.style={shape=circle,draw,minimum size=1.2em, fill=sron0, draw=sron0, text=white},
evil/.style={sron2, text=white},
channel/.style={very thick}, pay/.style={line width=1.5mm, rounded
corners=1mm, opacity=0.95, sron1, ->, >=stealth},
dots/.style={font=\footnotesize\sffamily, node distance=0.15cm}}

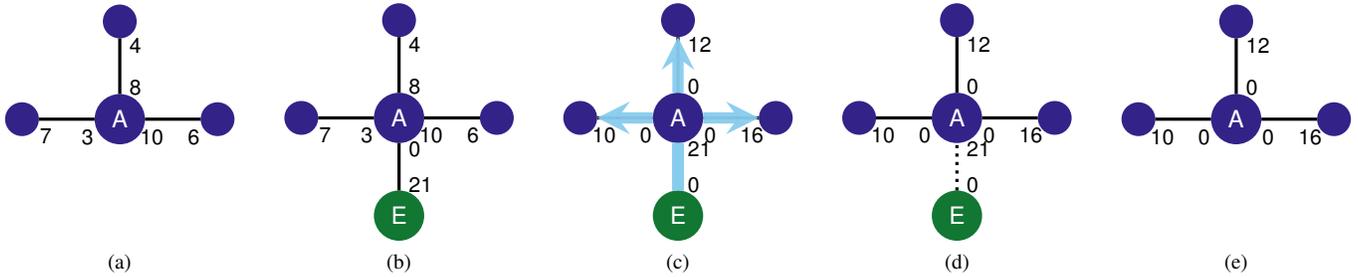
\begin{figure*}
    \centering
    \subfloat[]{%
        \centering
        \label{fig:isolationA}
        \begin{tikzpicture}[]
            \node[routenode]      	(a)      	at (0,0)                {A};
            \node[routenode, left of=a]      	(b)     			 		{};
            \node[routenode, above of=a]      	(c)     			 		{};
            \node[routenode, right of=a]      	(d)     			 		{};

            \node[routenode, evil, below of=a, draw=none, fill=none, text opacity=0]      	(e)     			 		{E};

            \draw[channel] (a)	-- 	(b)	node[font=\footnotesize\sffamily, very near start, below]	{3} node[font=\footnotesize\sffamily, very near end, below]	{7};
            \draw[channel] (a)	-- 	(c)	node[font=\footnotesize\sffamily, very near start,
            right]	{8} node[font=\footnotesize\sffamily, very near end, right]	{4};
            \draw[channel] (a)	-- 	(d)	node[font=\footnotesize\sffamily, very near start,
            below]	{10} node[font=\footnotesize\sffamily, very near end, below]	{6};

        \end{tikzpicture}
    }
    \hfill
    \subfloat[]{
        \centering
        \label{fig:isolationB}
        \begin{tikzpicture}[]
            \node[routenode]      	(a)      	at (0,0)                {A};
            \node[routenode, left of=a]      	(b)     			 		{};
            \node[routenode, above of=a]      	(c)     			 		{};
            \node[routenode, right of=a]      	(d)     			 		{};

            \node[routenode, evil, below of=a]      	(e)     			 		{E};

            \draw[channel] (a)	-- 	(b)	node[font=\footnotesize\sffamily, very near start, below]	{3} node[font=\footnotesize\sffamily, very near end, below]	{7};
            \draw[channel] (a)	-- 	(c)	node[font=\footnotesize\sffamily, very near start, right]	{8} node[font=\footnotesize\sffamily, very near end, right]	{4};
            \draw[channel] (a)	-- 	(d)	node[font=\footnotesize\sffamily, very near start, below]	{10} node[font=\footnotesize\sffamily, very near end, below]	{6};
            \draw[channel] (e)	-- 	(a) node[font=\footnotesize\sffamily, very near start, right] {21} node[font=\footnotesize\sffamily, very near end, right]	{0};

        \end{tikzpicture}
    }
    \hfill
    \subfloat[]{
        \centering
        \label{fig:isolationC}
        \begin{tikzpicture}[]
            \node[routenode]      	(a)      	at (0,0)                {A};
            \node[routenode, left of=a]      	(b)     			 		{};
            \node[routenode, above of=a]      	(c)     			 		{};
            \node[routenode, right of=a]      	(d)     			 		{};

            \node[routenode, evil, below of=a]      	(e)     			 		{E};

            \draw[channel] (a)	-- 	(b)	node[font=\footnotesize\sffamily, very near
            start, below]	{0} node[font=\footnotesize\sffamily, very near end, below]
            {10};
            \draw[channel] (a)	-- 	(c)	node[font=\footnotesize\sffamily, very near start,
            right]	{0} node[font=\footnotesize\sffamily, very near end, right]	{12};
            \draw[channel] (a)	-- 	(d)	node[font=\footnotesize\sffamily, very near start,
            below]	{0} node[font=\footnotesize\sffamily, very near end, below]	{16};
            \draw[channel] (e)	-- 	(a) node[font=\footnotesize\sffamily, very near start, right]	{0} node[font=\footnotesize\sffamily, very near end, right]	{21};
            \draw[channel, pay] (e)	-- 	(a) -- (b);
            \draw[channel, pay] (e)	-- 	(a) -- (c);
            \draw[channel, pay] (e)	-- 	(a) -- (d);

        \end{tikzpicture}
    }
    \hfill
    \subfloat[]{
        \centering
        \label{fig:isolationD}
        \begin{tikzpicture}[]
            \node[routenode]      	(a)      	at (0,0)                {A};
            \node[routenode, left of=a]      	(b)     			 		{};
            \node[routenode, above of=a]      	(c)     			 		{};
            \node[routenode, right of=a]      	(d)     			 		{};

            \node[routenode, evil, below of=a]      	(e)     			 		{E};

            \draw[channel] (a)	-- 	(b)	node[font=\footnotesize\sffamily, very near
            start, below]	{0} node[font=\footnotesize\sffamily, very near end, below]
            {10};
            \draw[channel] (a)	-- 	(c)	node[font=\footnotesize\sffamily, very near start,
            right]	{0} node[font=\footnotesize\sffamily, very near end, right]	{12};
            \draw[channel] (a)	-- 	(d)	node[font=\footnotesize\sffamily, very near start,
            below]	{0} node[font=\footnotesize\sffamily, very near end, below]	{16};
            \draw[channel, very thick, dotted] (e)	-- 	(a) node[font=\footnotesize\sffamily, very
            near start, right]	{0} node[font=\footnotesize\sffamily, very near end,
            right]	{21};
        \end{tikzpicture}
    }
    \hfill
    \subfloat[]{
        \centering
        \begin{tikzpicture}[]
            \node[routenode]      	(a)      	at (0,0)                {A};
            \node[routenode, left of=a]      	(b)     			 		{};
            \node[routenode, above of=a]      	(c)     			 		{};
            \node[routenode, right of=a]      	(d)     			 		{};

            \node[routenode, evil, below of=a, text opacity=0, draw=none,
            fill=none]      	(e)     			 		{E};

            \draw[channel] (a)	-- 	(b)	node[font=\footnotesize\sffamily, very near
            start, below]	{0} node[font=\footnotesize\sffamily, very near end, below]
            {10};
            \draw[channel] (a)	-- 	(c)	node[font=\footnotesize\sffamily, very near start,
            right]	{0} node[font=\footnotesize\sffamily, very near end, right]	{12};
            \draw[channel] (a)	-- 	(d)	node[font=\footnotesize\sffamily, very near start,
            below]	{0} node[font=\footnotesize\sffamily, very near end, below]	{16};
        \end{tikzpicture}
    }
    \caption{Node Isolation: the adversary $E$ establishes a sufficiently large
    payment channel to the target node $A$. It then exhausts all outgoing
    capacity of $A$ and closes her channel.}
    \label{fig:isolation}
\end{figure*}
\subsection{Attack Vectors}
\subsubsection{Denial of Service}

We consider denial-of-service (DoS) attacks as a general attack vector
to disrupt a node's connection to the Lightning Network by using \enquote{external} means,
\ie, not directly speaking the Lightning protocol.
DoS attacks are typically mounted by flooding nodes with superfluous requests
to overload their system.
We however also include a broader range of DoS attack techniques,
such as BGP hijacking to make nodes unreachable.

In general, a DoS attack on specific nodes in the Lightning Network allows an
adversary to inhibit these nodes from partaking in regular payment processing.
This attack vector usually requires a reasonably strong adversary
controlling a botnet or having access to the Internet backbone.
In March 2018, the Lightning Network was reportedly hit by a DDoS attack that
took 20\% of nodes offline~\cite{trustnodes2018ddos}. This
incident shows that DoS is not a mere theoretical threat,
but a feasible attack vector that has to be taken into account.

\subsubsection{Channel Exhaustion}
Each payment channel in the Lightning Network has a certain capacity and can
therefore only route payments up to this capacity.
We argue that this fact provides an attack vector:
an attacker with sufficient funds is able to exhaust,
\ie, block, a payment channel by routing a payment
over the targeted channel in the respective direction.
The attacker may not be able to infer the current channel balance
as nodes only announce a channel's initial capacity.
She therefore may need to route multiple payments
to eventually exhaust a channel.
To this end, the attacker could perform a binary search,
starting with the maximum channel capacity
and trying to route this volume.
In order to not waste funds,
the attacker is able to route funds back to herself,
in which case only marginal routing fees accrue.

Using this technique, an adversary is able to disturb
the payment flow in the network and manipulate it to its advantage. In
particular, this attack may be used to cut parts off the network graph,
leaving it in decomposed state.

\subsubsection{Payment Griefing}
The threat of channel exhaustion can currently be elevated
by combining it with an attack vector called
\emph{payment griefing}~\cite{robinson19htlcsharmful, griefing}:
as there is currently no fee on failed routing requests, the adversary may
initiate an arbitrary number of HTLC payments to a node under her control.
This node may then simply ignore incoming HTLC requests, forcing the involved
nodes to wait for the time locks to expire. Upon expiry, the entire
state is rolled back, circumventing fee deduction. Therefore, payment griefing
allows an adversary to temporarily claim channel capacity free of charge.

Channel exhaustion in general and payment griefing in particular
can be amplified by choosing longer payments paths.
In this case, the adversary's stake is able to exhaust/lock
funds along the path.
With payment griefing, though,
we can eliminate specific edges and paths in the PCN topology.

\subsubsection{Node Isolation}

By deliberately exhausting \emph{all} channels of a node,
we can isolate this node completely
and effectively hinder it from participating.
As shown in Figure~\ref{fig:isolation},
a malicious node $E$ can zero all outbound channel balances of a target node $A$.
The attack requires $E$ to first open a channel with a capacity
that is equal or greater than the total balance of $A$'s outbound channels (cf. Figure~\ref{fig:isolationB}).
Since the Lightning Network implements source routing,
the attacker is able to determine routes and exhaust each channel
by issuing a number of payments with respective payment volumes (cf. Figure~\ref{fig:isolationC}).
Of course, to improve efficiency, the attacker can also make use of payment griefing.
This attack vector can be considered a generalization
of the channel exhaustion, described previously.

Node isolation leaves the target node unable to route outbound payments.
While the node can of course still receive payment requests,
it is unable to fulfill them because all of its funds have been shifted
to the adversary's channel (cf. Figure~\ref{fig:isolationD}).
In this situation, both sides can decide to close the channel,
which would return the funds to the target node on-chain.
Note that this would additionally deter the target node
from using the funds for 1--2 block intervals, \ie, 10--20 minutes.
The adversary however could also leave the channel open
and refuse to process payment requests
at which point the target node is forced to close the channel unilaterally.
In this case, the settlement transaction can only be redeemed after the expiration a of the lock time,
which adds a delay that is typically larger than 1--2 block intervals.

This attack vector effectively incapacitates the node from functioning as a payment
hub, \ie, it eliminates the node from the routable network graph.
In order to recover from this state, the target node needs to open at least
one newly funded channel. As its old funds can only be reused 
after the closing channel was successful settled, it has to invest
additional funds to be able to start rebalancing its channels and eventually regain its routing capabilities.
Note that this comes with an additional overhead for the target node, since it
has to pay fees for the funding transaction in the Bitcoin network.

To conclude, even by design, it is possible to remove edges and nodes from the routable Lightning
Network. However, depending on the utilized attack vector, the adversary may
have to provide more or less resources to carry out the attack.

\subsection{Attack Strategies}
Equipped with the means to remove edges and nodes, the adversary may try to
cause maximum damage along the lines of the previously discussed adversary
goals. Depending on these goal, however, she may choose a different
strategy, \ie, a different set of nodes and edges to attack. In the following
we discuss a number of attack strategies.

\subsubsection{Highest Degree/Betweenness/Eigenvector Centrality Nodes}
An adversary aiming to damage the network as a
whole might try to destabilize or even partition the network graph, effectively
impeding oder hindering cross-network payment routing. For this, the adversary
might try to remove the participants according to their importance in the
network. Promising strategies therefore prioritize central nodes with respect
to some kind of centrality metric. Therefore, as an initial strategy, we propose that the attacker may target
nodes based on their \emph{degree} in descending order.

Moreover, the attacker may target nodes based on the previously mentioned
betweenness centrality~\cite{freeman1977set} or their eigenvector
centrality~\cite{gould1967geographical}, which can consider not only the
topological location, but also the edge capacities of a node.

\subsubsection{Highest Ranked Minimum Cut Sets}
A minimum cut set of a graph is a set of edges with minimal accumulated capacity that, when removed,
partitions the graph. Therefore, minimum cuts are prime target when an
adversary aims for network partitioning. However, not all cuts are created
equal: while some may partition the network quite effectively, others may only
cut off a single node.
Given that the adversary has only limited resources at her disposal, it is
important to prioritize the targeted minimum cuts according their importance
in real world payment scenarios. Therefore, we propose to calculate a high
number of potential $(s,t)$-cuts for randomly picked terminals $s$ and
$t$, and rank the individual cuts by the number of their occurrences. By
targeting the highest-ranked cuts, the adversary focuses on the network
bottlenecks hindering payment processing first.

\subsubsection{Highest Ranked Parallel Paths}
An adversary that participates in the network as a payment hub may be
interested in increasing her revenue by eliminating competitors. Of course,
the adversary may again target competing hubs by their importance in the
network, \eg, by node degree. However, such a strategy would not consider how
payments are routed in the Lightning Network. Therefore, we propose to
simulate random payments and record the resulting payment paths.
Excluding the paths involving the adversary's hub,
nodes can be ranked according to their involvement in the remaining payment paths.
Accordingly, the adversary eliminates nodes that are part of many competing routes
with the intention to increase her fees by processing more payments.

\subsection{Quantifying Adversarial Success}
In the following, we propose a number of metrics that allow us (and the
adversary for that matter) to quantify the impact of an attack strategy.
To provide an overall metric, we define the \emph{adversary's advantage} of an
adversary, \ie, success of the attack, as the \emph{relative decrease} in
the respective metric concerning the a priori measurement~$m$ and the a
posteriori measurement~$m'$:
\begin{equation*}
    \Delta_m = \left|\frac{m-m'}{m} \right| = \left| 1 - \frac{m'}{m} \right|.
\end{equation*}
The higher $\Delta_m$ becomes the higher the adversary's success according to
metric $m$ will be.
It generally provides a way to relatively compare the prospect of different attack strategies
from different perspectives.

An attacker may try to partition the network into a number of subgraphs.
While the impact of such an attack is limited when the adversary is only
excluding single nodes, it may be much more severe if she can cut off larger
segments of the network. A sound measure for general network robustness should
capture the share of nodes that are disconnected from the network graph.
We therefore propose the \emph{number of reachable nodes} $r$ as a metric.
Given all connected components $C_i$, we define the largest connected
component~$C_1$ to be \emph{the} Lightning Network.
Accordingly, we can calculate $r$ as the network's node cardinality $r = |C_1|$.
This metric can be used to calculate the adversarial advantage~$\Delta_r$
as defined above.

However, as some nodes are more central and provide a larger share of the
network's total capacity than others, the impact of node isolation on the
liquidity of the network may vary heavily depending on the target. To quantify
the impact, we propose the \emph{average maximum flow} as another metric:
for $n$ rounds, we draw a pair of nodes~$s_i,t_i \in V,\, i \in \{1\ldots n\}$
by uniform random sampling and calculate the maximum flow~$F_i(s_i,t_i)$
along the lines of~\cite{ford1956maximal, goldberg1988new}.
The average maximum flow is then given by
\begin{equation*}
    \overline{F} = \frac{\sum\limits_{i = 1}^{n} F_i(s_i, t_i)}{n}
\end{equation*}
and can be used to calculate the adversarial advantage~$\Delta_F$.

While the average maximum flow is a good indicator of the routable capacity in
the network, it does not necessarily reflect the actual expected payment
success, since currently the Lightning Network only uses single-path routing
to fulfill payments.
Therefore, we additionally introduce the \emph{expected payment
success ratio}~$s$ as a metric for how likely payments can be processed by the network.
To get a sound estimation, we simulate a high number of transactions between random nodes
and calculate the success ratio as
\begin{equation*}
    s = \frac{\#\text{successful payments}}{\#\text{attempts}}.
\end{equation*}
Accordingly, the adversarial advantage is given by~$\Delta_s$.
As the validity of this measure heavily depends on the transaction model,
it will especially benefit from parametrization based on empirical data.

In order to quantify the potential success of an internal adversary aiming for increased
revenue, we propose to simulate a high number of payments and accumulate the
fee gain $g_{i,h}t$ for the adversary's hub $h_a$ that accrues over
all simulated payments $i \in \{1 \ldots n\}$. The \emph{average fee gain}
\begin{equation*}
    \overline{g_h} = \frac{\sum\limits_{i = 1}^{n} g_{i,h_a}}{n}
\end{equation*}
may then be used to indicate the adversary's success.

\section{Evaluation}\label{sec:evaluation}

\subsection{Proof of Concept}
In order to validate the feasibility of our node isolation attack, we built a
simple toy scenario mimicking the attack shown in Figure~\ref{fig:isolation}.
We ran five independent \texttt{lnd} instances,
which were connected to the Bitcoin testnet.
The target node~$A$ established three channels with outbound capacities set to
75,000, 100,000, and 125,000 satoshis, respectively.
The attacker $E$ established a channel with a total capacity of 400,000 satoshis to $A$,
which is sufficient to exhaust $A$'s channels and therefore hinder any other node
from routing through $A$. In our example, we repeatedly sent payments of declining
size until $A$ was able to route no more than $100$ satoshis (currently $\approx
0.0037$ USD), at which point we considered the attack to be
successful.

\subsection{Evaluation Model}
The following evaluation of topology-based attacks on the Lightning Network is
based on simulations we implemented using
\texttt{networkx}~\cite{hagberg2008exploring}.
As before, the snapshot from Feb.~1,\ 2019 is used as our reference dataset.
While the dataset provides real-world data on nodes,
edges, and edge capacities, it does not include the actual channel balances.
We therefore assumed the given capacities to be the balance both ways,
likely resulting in an overestimation of the routable funds, \ie, yielding a
best-case estimation for the considered metrics.

The proposed metrics and attack strategies rely on the availability of a solid
payment model that reflects how transactions of a certain volume traverse the
network. Due to the lack of real-world transaction data of the Lightning Network
we draw source and
target nodes uniformly at random from the network nodes.
By doing this, we refrain from introducing unnecessary complex and artificial assumptions.
We also assume a
single-path routing scheme as currently implemented by the Lightning Network:
each payment is processed by first excluding all edges with insufficient
capacities from the routable network graph.
On the remaining graph, shortest path routing is performed.

As our data base for payment volumes, we collected real world payment data from the Ripple
network~\cite{rippledata}. For this, we retrieved all XRP transactions that
occurred at our reference date Feb.~1,\ 2019 and converted it to the
respective values in satoshis. The
transaction volumes are chosen by uniform random sampling from this data set.

All algorithms are repeated 1,000 times to ensure statistical significance.
Furthermore, to ensure the reproducibility of the applied metrics, we opted to
fixate the pseudorandom number generator's seed value
for each round of simulation. Thereby, the same input data is utilized by all
metrics, improving the comparability between measurements.

\subsection{Partitioning Attacks}
As we have seen, a capable adversary may isolate single nodes in the network.
In the following, we analyze the Lightning Network's resilience to an adversary
aiming for maximal damage to the network, \ie, network partitioning.
To this end, we assume that the attacker is capable of removing a certain number
of nodes from the routable network graph, \eg,
by the means of DoS attacks or node isolation attacks.
We simulated the previously introduced attack strategies
and recorded the network state before (a priori) and after (a posteriori) the attack:
removing nodes by decreasing \emph{degree}, by
decreasing \emph{betweenness} and \emph{eigenvector} centrality, and by
highest ranked minimum \emph{cuts}. Moreover, we evaluated uniform
\emph{random} node removal as a baseline.

\begin{figure}
    \input{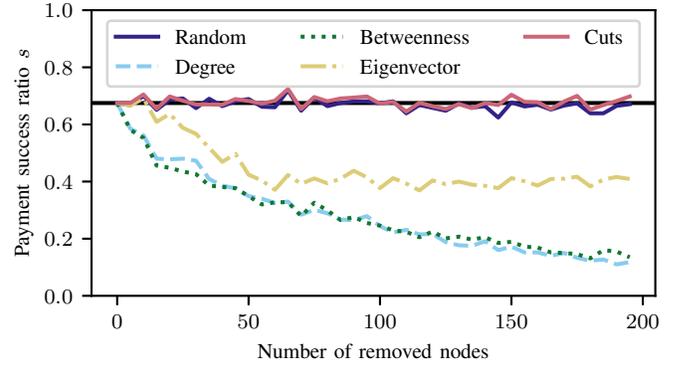}
    \caption{Payment success ratio $s$ before (horizontal line) and after the
    removal of $n$ nodes according to different attack strategies.}
    \label{fig:success_abs}
\end{figure}
\begin{figure}
    \input{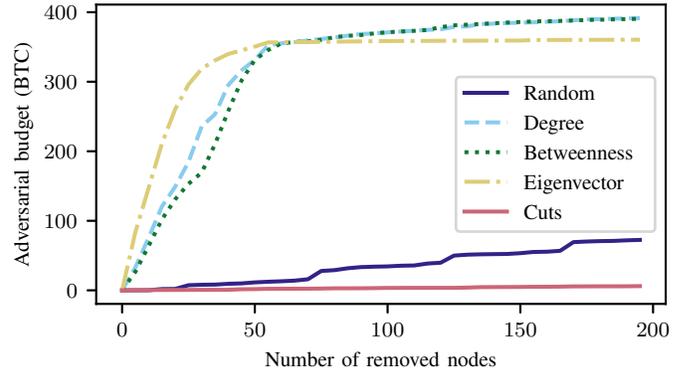}
    \caption{Required adversarial budget to remove $n$ nodes by node isolation
    and according to
    different attack strategies.}
    \label{fig:spents}
\end{figure}

\begin{figure*}
    \input{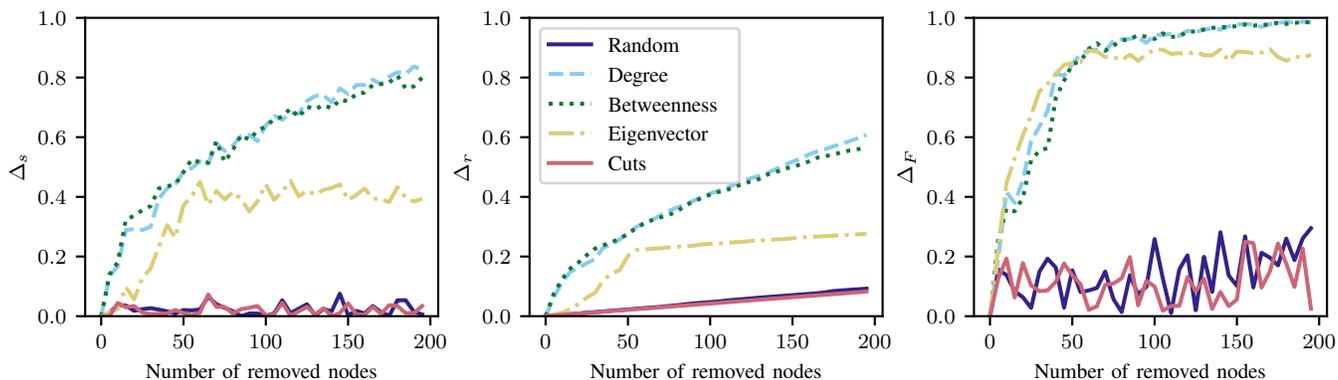}
    \caption{Adversary's advantage for success ratio, reachability, and
    average maximum flow after the removal of $n$ nodes according to different
    attack strategies.}
    \label{fig:three}
\end{figure*}

\begin{figure*}
    \input{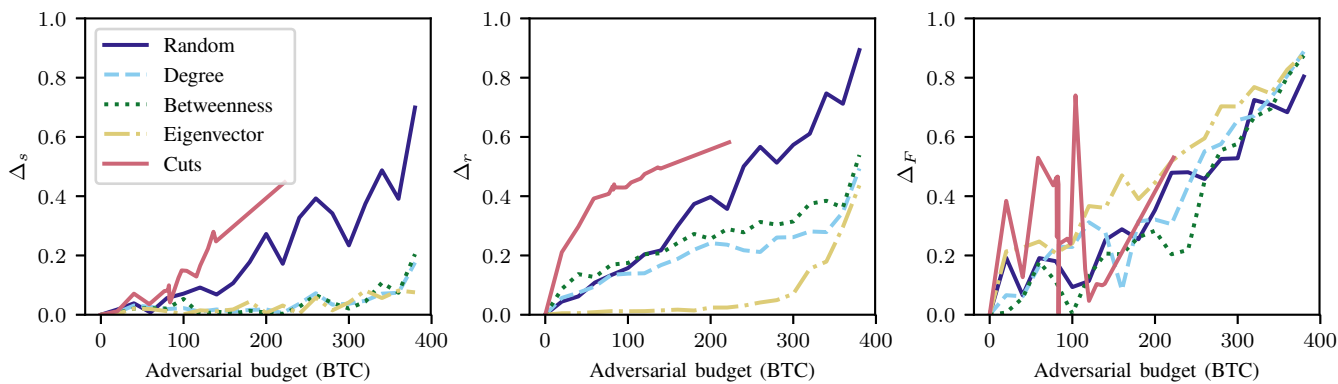}
    \caption{Attainable adversary's advantage depending on her budget for node
    isolation attacks.}
    \label{fig:three_budget}
\end{figure*}

As an initial measure, Figure~\ref{fig:success_abs} shows the payment success
ratio $s$ before (horizontal line) and after the respective attacks happened.
Notably, even before the attack only $675$ out of the 1,000 tried payments
succeeded. Moreover, we can see that the strategies work out quite
differently, with degree-based removal and betweenness-based removal resulting
in a steep decrease of the success ratio (down to around 10\% success), while the random and minimum cut
strategies measure around the baseline.
Accordingly, the adversary's advantage
for the success rate $\Delta_s$ is the highest for the former strategies, as
shown in Figure~\ref{fig:three}. Similar results can be seen for the
reachability, where the degree and betweenness strategies achieve $\Delta_r$
values of larger than $0.5$, \ie, cutting off more than half of the network.

Moreover, the average maximum flow of the network is heavily impaired when
removing central nodes. Here, the eigenvector strategy is superior until it is
outperformed by the degree and betweenness strategies.
The latter two eventually lead to a near-total collapse of the
maximum flow, \ie, $\Delta_F \approx 1.0$, rendering the remaining network useless.

So far, targeting nodes based on their centrality seems to be the most
promising strategy for an adversary that is capable of eliminating a certain
number of nodes. %
However, as we see in Figure~\ref{fig:spents},
the centrality-based strategies also require the largest budget
to successfully mount node isolation attacks, while the minimum cut strategy
exhibits really low budget requirements.
In the following, we therefore evaluate the efficiency of our attack strategies.

\subsection{Efficiency of Node Isolation Attacks}
In order to evaluate the efficiency of the attack strategies, we assigned each
adversary a budget and then analyzed the \enquote{damage} it can cause.
In particular, we simulate a node isolation only,
if the attacker's (remaining) budget is large enough to remove all edges.
Otherwise, the simulation skips the node
and tries to utilize the available funds on the next target proposed by the respective strategy.
Likewise, we only remove complete cuts. As a consequence, the minimum cut
strategy does not always consume the full budget. 

The results are shown in Figure~\ref{fig:three_budget}:
independently of the strategy, we observe that it requires high budgets to
give the adversary the power to reliably disturb all payment attempts,
which would result in a high advantage score $\Delta_s$. Notably, the
previously underperforming minimum cuts and random node removal strategies
exhibit the best efficiency properties, \eg, an adversary could attain an
advantage of $\Delta_s = 0.4$ when spending around $200$ BTC.
Similar behaviour can be seen for the impact on the adversary's advantage in terms of the reachability~$\Delta_r$
and average maximum flow~$\Delta_F$. While some
strategies seem to be subject to fluctuations, which do not always allow to infer a
clear ordering in efficiency, the highest ranked minimum cut strategy again clearly stands out as the most
efficient in terms of the $\Delta_r$ and $\Delta_F$,
exhibiting values above $0.5$ in both cases. This is
not surprising, since the maximum cut strategy targets the connecting edges and nodes first, whose
removal has a significant impact on graph connectivity and the available
network capacity.

\subsection{Fee Gain}
In contrast to disrupting the network,
an adversary might be interested in increasing its own profit
by strategically eliminating competing nodes.
We assumed that this adversary is an established payment hub in the network
(amongst the top 10 nodes ranked by total capacity).
According to our strategies,
most notable highest ranked degree and highest ranked parallel paths,
we eliminated up to 30 nodes.
We used the average fee gain as metric to quantify the adversary's success.
While the adversary can indeed profit from eliminating nodes,
we cannot observe a clear trend.
In fact, we believe that node-based elimination strategies are too coarsely grained.
Instead, we argue that an channel-based elimination strategy might be superior,
which we intend to investigate in the future.

\subsection{Discussion}
In our analysis, we assume that each node functions as a payment hub, \ie,
accepts incoming payment channels. Moreover, the attacker is assumed to find
an adequate endpoint with sufficient capacity to route payments.
This is especially important if the attacker does not
execute payment griefing attacks, but needs to route payments back to herself.
In this case, the attacker has to ensure that some nodes in the network
first establish channels of sufficient volume to her secondary node,
\ie, her target node is connected with high enough inbound capacity.

In order to mitigate the possibility of node isolation attacks, the client
software should employ rate limiting techniques to limit the number of
incoming channels and incoming channel volume. This would make it harder for
an adversary to quickly establish high-volume channels from an advantageous position in the
topology. However, a client probably cannot mitigate the risk of node
isolation attacks entirely, since the attacker may circumvent simple rate
limiting strategies by splitting the funds over multiple identities and channels.

Moreover, network partitioning attacks may be counteracted by the
so-called \emph{autopilot} algorithms responsible for automated payment channel creation.
This may be achieved by monitoring previously discussed metrics and
restructuring the topology accordingly to make it less susceptible to targeted attacks.

\section{Related Work}\label{sec:relwork}

Payment channels were introduced to scale cryptocurrencies to high transactions rates.
Several channel designs have been proposed over the past
years~\cite{miller17sprites,green17bolt,wattenhofer2015duplexmicropayment,lind16teechan,heilman17tumblebit}.
While some designs are restricted to single-hop payments,
others can also be used for multi-hop payments and therefore are qualified for PCNs.
The Lightning Network for Bitcoin and the Raiden Network for Ethereum
emerged as the most prominent PCN implementations.
A technical overview focusing on Bitcoin's PCN design space is provided by~\cite{mccorry16pcn}.

Most of the research in the area focuses on challenges
concerning privacy, concurrency, and
routing~\cite{prihodko2016flare,malavolta17pcn,rohrer17towards,werman18pcnDeadlocks,roos2018settling}.
Concurrently to our work,
the authors of~\cite{seres19lnanalysis} explore the Lightning Network's topology
and also confirm that the network exhibits properties similar to small-world and scale-free networks.
Security of PCNs was mainly discussed in the context of dispute handling~\cite{mccorry18pisa,avarikioti18watchtower},
which assumes each party to be responsive and synchronized with the blockchain at all times.
While this discussion relates to node failures as well,
security in general and topology-based attacks in particular
have been mostly neglected, though.

When it comes to the mainnets of major cryptocurrencies, most notably Bitcoin and Ethereum,
network and centrality measurement studies~\cite{gencer2018decentralization,miller2015discovering,baumann2014exploring}
as well as security analysis concerning node isolation~\cite{heilman2015eclipsebitcoin} exist.
Beyond cryptocurrencies, many topology analyses have been conducted.
For example, the peer-to-peer network Gnutella has been classified as a small-world network~\cite{jovanovic2001modeling}.

In general, our work is orthogonal to the previous work
and therefore contributes an important new perspective
to the area of PCNs.

\section{Conclusion}\label{sec:conclusion}
In this paper, we analyzed the Lightning Network's payment channel topology
and investigated its resilience towards random failures and targeted attacks.
We've shown that the current Lightning Network can be subjected to
\emph{channel exhaustion} or \emph{node isolation} attacks and that these
attack vectors may indeed have severe consequences for the payment success
ratio, reachability, and average payment flow of the network.

\balance
\printbibliography[notcategory=skipbib]
\end{document}